# Direct observation of the surface superconducting gap in the topological superconductor candidate β-PdBi$_2$

Akifumi Mine[1], Takeshi Suzuki[1,*,†], Yigui Zhong[1], Sahand Najafzadeh[1], Kenjiro Okawa[2,‡], Masato Sakano[3,§], Kyoko Ishizaka[3,4,5], Shik Shin[1,6,7], Takao Sasagawa[2,8], and Kozo Okazaki[1,7,9,¶]

[1]Institute for Solid State Physics, University of Tokyo, Kashiwa, Chiba 277-8581, Japan
[2]Materials and Structures Laboratory, Institute of Science Tokyo, Yokohama, Kanagawa 226-8501, Japan
[3]Department of Applied Physics, The University of Tokyo, Hongo, Tokyo 113-8656, Japan
[4]Quantum-Phase Electronics Center (QPEC), The University of Tokyo, Hongo, Tokyo 113-8656, Japan
[5]RIKEN Center for Emergent Matter Science (CEMS), Wako, Saitama 351-0198, Japan
[6]Office of University Professor, University of Tokyo, Kashiwa, Chiba 277-8581, Japan
[7]Material Innovation Research Center, University of Tokyo, Kashiwa, Chiba 277-8581, Japan
[8]Research Center for Autonomous Systems Materialogy, Institute of Science Tokyo, Yokohama, Kanagawa 226-8501, Japan
[9]Trans-scale Quantum Science Institute, University of Tokyo, Bunkyo-ku, Tokyo 113-0033, Japan

β-PdBi$_2$ is one of the candidates for topological superconductors with a superconducting (SC) transition temperature ($T_c$) of 5.3 K, in which parity mixing of spin singlet and spin triplet has been anticipated, being crucial for the further understanding of relationship with inversion symmetry and parity mixing in the superconductivity. In this work, we measured the SC gap in high-quality single crystal of β-PdBi$_2$ by using high-resolution laser angle-resolved photoemission spectroscopy below $T_c$. We found the isotropic SC gaps in momentum space for multiple bands, and observed that the difference between the SC gap of the topological surface bands and the bulk bands is about 0.1 meV, consistent with other experimental results. These direct and quantitative experimental results support the possibility of β-PdBi$_2$ as a topological superconductor, characterized by unique crystal and electronic band structures.

Topological materials have emerged as a central theme in condensed matter physics, with topological insulators (TIs) being the first extensively studied example [1,2]. TIs are characterized by an insulating bulk gap and spin-polarized Dirac-cone-like surface states protected by time-reversal symmetry [1,2]. The concept of band topology has since been extended beyond insulators to superconductors [3,4] and semimetals [5,6], giving rise to diverse topological classes distinguished by their crystal symmetry, dimensionality, and quasiparticle statistics [7]. Among them, topological superconductors (TSCs), which host superconducting states with nontrivial topology, are of particular interest due to their potential to realize Majorana quasiparticles [8], offering a promising platform for topological quantum computation [9,10]. One promising route to TSCs is to induce superconductivity in topological surface states via proximity effects [11,12] or chemical doping [13].

However, experimentally confirmed candidates remain limited, and the unambiguous realization of intrinsic TSCs continues to be a subject of active debate [3,4,9,14]. For example, non-centrosymmetric α-BiPd is one of the few superconductors with topological surface states [15,16]. However, recent experimental studies suggest that its superconducting state is topologically trivial, exhibiting an s-wave-like SC gap with negligible contributions from spin-triplet pairing [15]. Another notable candidate is PbTaSe$_2$. However, similar to α-BiPd, it has a low $T_c$, which limits its suitability for ARPES measurements [17,18]. Furthermore, the topological nodal-line states in PbTaSe$_2$ lie above the Fermi level ($E_F$), making it difficult to directly observe the SC gap associated with these states, which is crucial for understanding their physical properties [17,18].

Trigonal PtBi$_2$ exhibits a large SC gap in the Fermi arc, as observed by ARPES. Interestingly, the surface state exhibits superconductivity up to 5 K, while the bulk $T_c$ is about 1 K. However, due to the low $T_c$ of the bulk, it is difficult to compare the SC gap magnitudes of the surface and bulk states using ARPES. Fe(Se,Te) has been also demonstrated as a topological superconductor by ARPES, where an isotropic SC gap hosted from the degenerate bulk bands is directly observed in the surface states [21]. However, the surface superconductivity in Fe(Se,Te) is well separated from the bulk bands in momentum space, and only a weak proximity effect works [21]. In addition, the topological state is realized by chemical substitution, inducing inhomogeneity, and thus it hinders reproducibility in applied field [22].

β-PdBi$_2$ has been proposed as another candidate for a topological superconductor with a SC transition temperature ($T_c$) of 5.3 K. The previous ARPES measurements with spin resolution reported the existence of topological surface states with a helical spin structure [23]. The significant feature of β-PdBi$_2$ is its centrosymmetric crystal structure that can preserve inversion symmetry in the bulk, which quantitatively differs from PbTaSe$_2$ with a non-centrosymmetric crystal structure [17,23]. Although the inversion symmetry of the surface state is broken, that of the bulk state is maintained, which has been experimentally confirmed [23,24]. Furthermore, relativistic first-principle calculations pointed out that the surface states of β-PdBi$_2$ have nondegenerate spin-polarized because the bulk states are topologically nontrivial [23,24]. Superconductivity arising from nondegenerate spin-polarized Fermi surfaces is expected to consist of a mixture of spin singlet and triplet for inversion symmetry breaking [24-26]. If superconductivity is induced in spin-polarized topological surface states, the induced spin-triplet order parameter will play an important role in the appearance of Majorana fermions in edge states and vortex cores [24,27,28]. In addition, theoretical calculations and ARPES measurements show that the bulk and surface bands are adjacent in momentum space and that the proximity effect is significant, making the evaluation of the superconducting pairing noteworthy [23].

In order to prove that β-PdBi$_2$ is a topological superconductor, it is crucial to directly observe the SC gap in the topological surface states in momentum space. In addition, our goal of this study is to verify parity hybridization by comparing the SC gaps in the surface states with broken inversion symmetry suggested and in the bulk states with inversion symmetry. However, β-PdBi$_2$ has a substantially low $T_c$ and its SC gap is expected to be very small, especially in topological surface states, which limits experimental techniques to access their SC gaps in momentum space.

In this study, we have investigated the electronic structure of β-PdBi$_2$ in the superconducting state to confirm whether β-PdBi$_2$ is a topological superconductor or not by using low-temperature and high-resolution laser-based angle-resolved photoemission spectroscopy (laser ARPES) [29]. We successfully observe the SC gap of the topological surface states as well as the bulk states. Furthermore, the SC gap size is found to be isotropic in momentum space in each Fermi Surface (FS). The difference of SC gap size between the surface and bulk bands is found to be about 0.1 meV, which is consistent with the STM results. We can attribute this difference not only to proximity effect, but also to the parity mixing due to the breaking of inversion symmetry in the surface states, where the bulk band consists only of the spin-singlet band while the surface band is influenced by both the spin-singlet and triplet states.

β-PdBi$_2$ has a centrosymmetric tetragonal layered crystal structure belonging to the space group I4/mmm shown in Fig. 1(a). Each Pd atom is located at the center of a square prism consisting of eight Bi atoms, and each layer can be easily exfoliated owing to van der Waals stacking [23,24,30]. Figure 1(b) shows the schematic band structure of β-PdBi$_2$ near $E_F$. The bands are called α band, β band, and γ band in order from the Γ-M point, and there is a δ band around X point. These four bands are bulk bands, and there are two surface bands in Γ-M direction, which are called S1 band and S2 band, adjacent to β and γ bands in momentum space [23]. Regarding these two surface bands, while S2 is topologically nontrivial, S1 is topologically trivial [23].

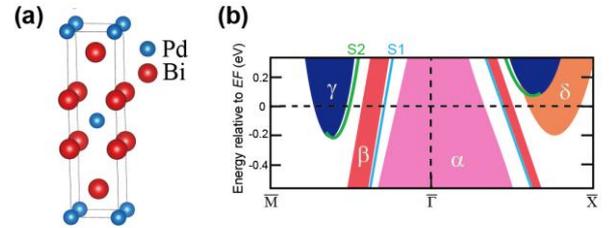

FIG.1. (a) Crystal structure of superconductor β-PdBi$_2$. *a*, *b* and *c* axes are taken along the body-centered tetragonal crystal orientation. (b) Schematic band structure around $E_F$. It consists of four bulk bands and two surface bands.

Measurements were performed with a laser-based ARPES setup, which consisted of a true continuous-wave laser (h$\upsilon$ = 5.8 eV) and a quasi-continuous-wave laser (h$\upsilon$ = 6.994 ≃ 7eV) with a repetition rate = 960 MHz, a Scienta HR8000 hemispherical analyzer, and a sample manipulator cooled by decompression-evaporative cooling of liquid helium [29]. The samples were cleaved in situ and measured under an ultrahigh vacuum better than $3×10^{-11}$ torr. The sample temperature varied from 2 to 10 K, and the energy resolution for the SC gap measurements was less than 0.1 meV (1σ < 0.03 meV) for 5.8-eV laser and 0.3 meV (1σ < 0.1 meV) for 7eV laser. The resolution is determined by the stability of the Fermi level at the same angle sap, the margin of error of the slight Fermi level that occurs in multiple $E_F$ measurements [31]. After checking the linearity of the detector, the Fermi level was calibrated with the EDC fit at the same angle in the normal state and an in-situ connected gold and high temperature sample reference. Crystals of β-PdBi$_2$ were grown by a melt growth method. The obtained single crystals had good cleavage, producing flat surfaces as large as ~1 × 1 cm$^2$. This sample was fabricated using the same method as in other papers [23,24]. Resistivity and magnetic susceptibility exhibit a clear superconducting transition at $T_c$ = 5.3 K.

Figure 2(a) shows the FS map measured by 7 eV laser at

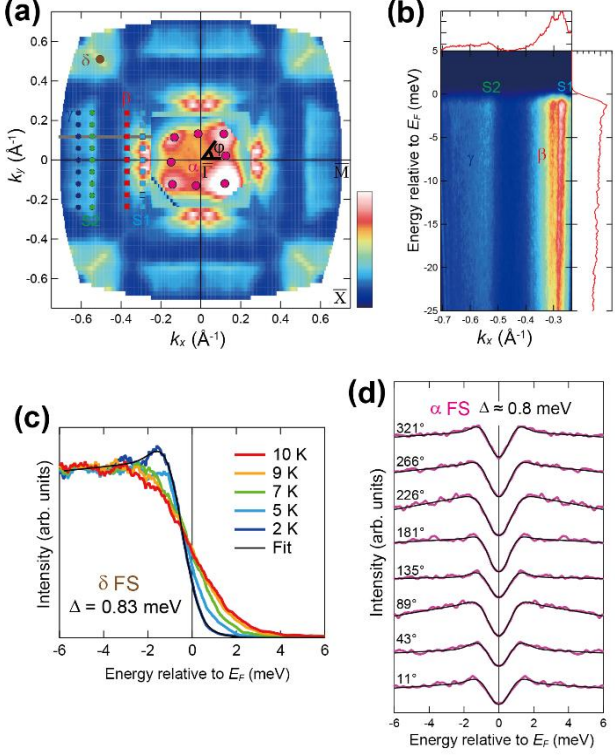

FIG.2. (a) Fermi surface map measured by a 7 eV laser. Due to the polarization dependence of each FS, the mapping of α FS is measured with S polarization, while the symmetrized mappings of β, S1, γ, S2, and δ are measured with P polarization. (b) E-k map at $k_y$ = 0.05 (1/Å), and EDC and MDC along EF and $k_x$ = 0.35 (1/Å). The broad band around $k_x$ = -0.65 (1/Å) is the γ FS, and the S2 FS is next to γ band around $k_x$ = -0.55 (1/Å). The β FS is near $k_x$ = -0.3 (1/Å), and the most intense band near $k_x$ = -0.33 (1/Å) is the S1 FS. (c) Temperature dependent EDCs of the brown point in the δ FS. (d) SC gaps in the α FS. Each symmetrized EDC cut is corresponding to the angle points in the Fig.2(a).

2 K. It is combined the results for the symmetrized Fermi surface near the M point with the results for the Γ point because it is not possible to measure the entire Fermi surface simultaneously due to limitations in the measurable momentum space [31]. Figure 2(b) shows the E-k map along the gray line in Fig.2(a). These obtained FS map and band dispersion are consistent with the previous results of the synchrotron ARPES [23]. The S1 and S2 bands, which are clearly observed as surface bands near the $E_F$, play a major role in the topological nature of β-PdBi$_2$.

Now we focus on the SC gap of β-PdBi$_2$. First, Fig. 2(c) shows the temperature dependence of energy distribution curves (EDCs) at the brown point in δ FS. The coherence peak and the leading-edge shift are observed at 2 K, whereas no gap is observed in the $E_F$ at higher temperatures. Also, the black line in Fig. 2c shows the fitting results, and the evaluated gap size is about 0.83 meV. To estimate the size of the SC gap, we performed the fitting by using a Bardeen-Cooper-Schrieffer (BCS) spectral function [29,33]. Also, Fig. 2(d) shows the symmetrized EDCs of the α FS corresponding to the markers indicated in Fig. 2a. The SC-gap size is isotropic in momentum space and approximately 0.8 meV.

Next, Figs. 3(a)-(d) show the symmetrized EDCs of γ, S2, β, and S1 FSs at 2 K. The measured positions for each panel in momentum space are denoted in Fig. 2(a). Figure 3(e) is a summary of the momentum-dependent SC gap for γ, S2, β, and S1 FSs as a function of $k_y$. The size of the SC gap Δ in the β and γ FSs ranges from 0.79 to 0.82 meV suggesting an almost isotropic nature in momentum space. For the S2 and S1 FSs, Δ ranges from 0.67 to 0.71 meV, also showing isotropic in momentum space. According to these results, all the measured bulk-derived α, β, γ, and δ FSs have isotropic SC gap structures with a gap size of approximately 0.8 meV. In addition, these results suggest that the surface-band-derived FSs, i.e., S1 and S2, have approximately 0.1 meV smaller SC gaps than the bulk-band-derived FSs, i.e., β and γ. The energy resolution of 7 eV laser is about 0.3 meV, which is larger than 0.1 meV, but, considering the tolerance range of σ, the fact that multiple EDCs show this result indicates that the 0.1 meV difference in the SC gap between bulk-derived and surface-derived FS is significant.

To confirm the above finding, we performed the measurements with a 5.8 eV continuous-wave (CW) laser around the Γ point. Figure 3f shows the FS map measured by using the 5.8 eV laser. Using the 5.8 eV CW laser, we can perform the measurements with an energy resolution of 0.1 meV, which is better than 7 eV laser, because the 7 eV laser is quasi-continuous wave, increasing the intensity leads to unavoidable space charge effects, whereas the 5.8 eV laser, being continuous wave, can avoid space charge effects even when the intensity is increased. On the other hand, the 5.8 eV CW laser, due to its lower photon energy, can only measure a limited momentum range near the Γ point. As a result, while the S2 band is accessible, this represents the limit of the 5.8 eV laser, and unfortunately, the γ band cannot be reached, even when the sample is tilted to its maximum angle. The symmetrized EDCs for the β and S1 bands obtained with the 5.8 eV laser are shown in Fig. 3g, which reveal a distinctive difference between them, thus confirming the consistency between the gap sizes estimated by 7 eV and 5.8 eV lasers. The momentum positions of each symmetrized EDC correspond to those denoted in Fig.3(f). While the bulk-derived β FS shows the SC gap of 0.78 meV, the surface-derived S1 FS shows the SC gap of 0.69 meV. These results are consistent with the results obtained with the 7 eV laser that the SC gap of the surface-band-derived FS is about 0.1 meV smaller than that of the bulk-derived FS.

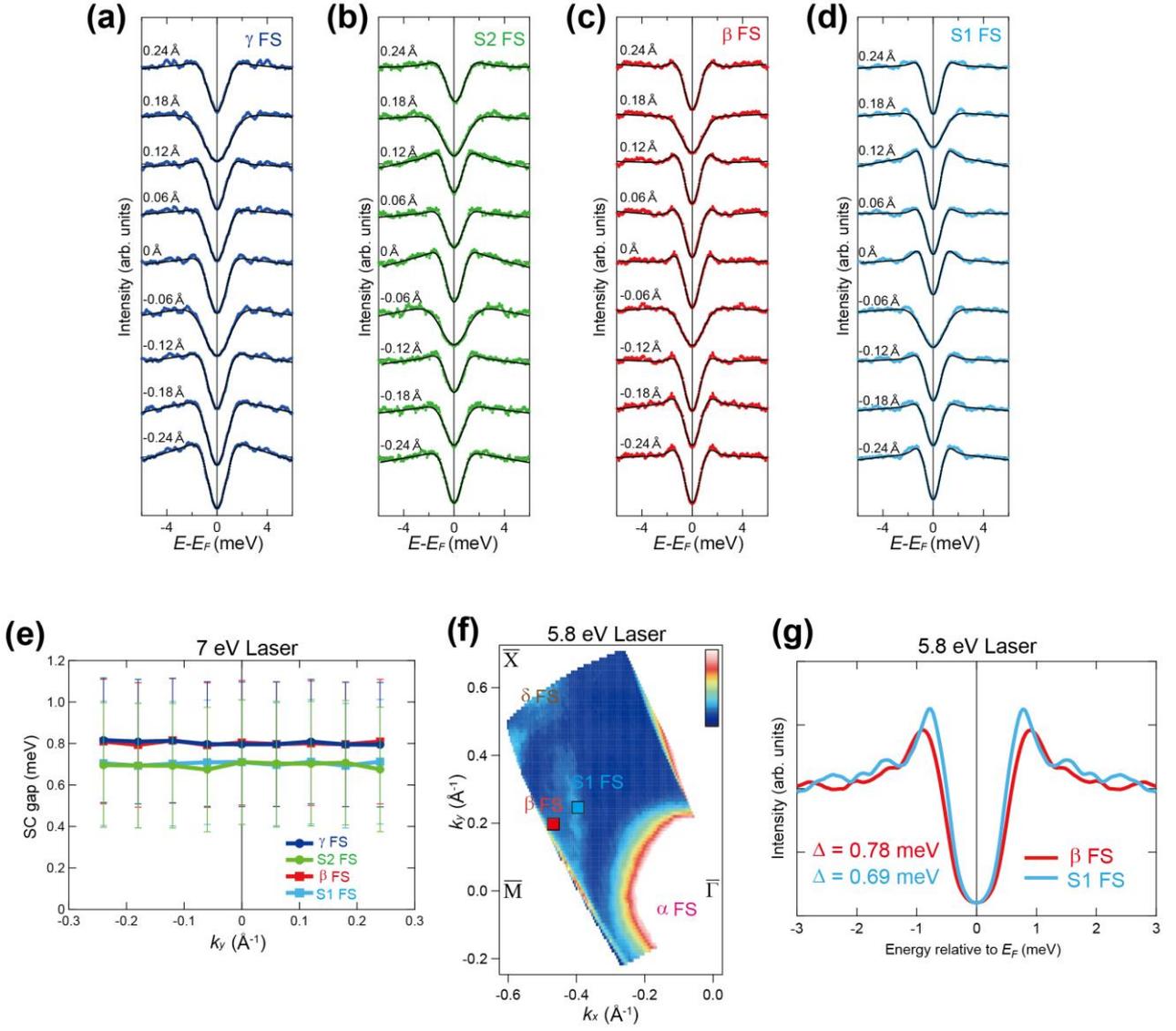

FIG. 3. (a-d) SC gaps in γ, S2, β, and S1 FSs. Each symmetrized EDC cut is corresponding to the plot to the $k_y$ momentum in Fig.2(a). (e) The plot of the SC-gap size in γ, S2, β, and S1 FSs. (f) Fermi surface map measured by 5.8 eV laser around Γ point. (g) SC gaps in the β, and S1 FS. Each symmetrized EDC cut is corresponding to the points in the Fig.3(f).

Also, the momentum perpendicular to the cleaved surface $k_z$ are assumed to be different between the 5.8 eV and 7 eV lasers. Comparing the gap sizes of the β FS observed with 7 eV and 5.8 eV lasers shown in Figs. 3(c) and 3(g), we see that they are very close values of 0.8 meV, indicating that the SC gap of the β FS is almost isotropic in the $k_z$ direction.

However, since we only performed the measurements with two photon energies, there is still room for further investigation of the $k_z$ dependence of the SC gap.

This study has revealed several important points. First, we have substantiated that β-PdBi$_2$ could be a topological superconductor by measuring the SC gap in the topological surface bands, in a similar fashion to Fe(Se,Te) [21,23]. An interesting feature of β-PdBi$_2$ is that the surface-derived bands are adjacent to the bulk-derived bands in momentum space [23]. This allows us to evaluate the effect of inversion symmetry breaking on the SC gap by comparing the SC gap in the surface-derived band with that in the adjacent bulk-derived band. Since the inversion symmetry is broken at the surface, spin-singlet and -triplet states can mix, resulting in a parity mixing state [24]. Thus, the difference between the

SC gap amplitude of the bulk- and surface-derived bands implies the contribution from the spin triplet component [24,34,35]. Theoretical calculations predict that the SC gap of the surface-derived band is comparable to or slightly smaller than that of the adjacent bulk-derived band [24,36], which is consistent with our experimental result that the gaps of surface-derived bands are about 0.1 meV smaller than that of the bulk-derived bands.

As for the SC paring strength of β-PdBi$_2$, the SC gap size of the bulk-derived bands is about 0.8 meV, corresponding to $2\Delta/k_BT_c$ = 3.6 almost the same value of the BCS theory ($2\Delta/k_BT_c$ = 3.53). The superconducting pairing in the surface state is a littile bit weaker than the bulk, $2\Delta/k_BT_c$ = 3.2. The surface superconductivity is induced by the proximity effect from the adjacent bulk superconductivity; thus the SC paring strength of the surface sates is more or less weaker than that of the bulk states [37]. The magnitude of this effect increases more significantly when the surface- and bulk-derived bands locate closer in momentum space, and the effect of parity mixing can also be relevant. When the bulk and surface bands become close to each other in momentum space, the bulk state hybridizes with the surface state and the spin triplet component is suppressed. On the other hand, if the two bands are too far apart, it is expected the proximity effect weakens and superconductivity is no longer induced in the surface state. Therefore, it is important to strike a balance between these two effects. In the case of β-PdBi$_2$, the proximity effect is strong enough and the parity mixing effect is probably finite because the surface and bulk bands are so close together. Thus, both the proximity effect and the parity mixing must be considered to account for the difference between the surface and bulk SC gaps. However, it is difficult to completely distinguish between these two effects with our method, and this is an issue for future work.

The isotropic SC gap in the bulk- and surface-derived bands observed in this study is consistent with other experiments [24,38]. These results support the dominance of the s-wave component in the superconducting ground state of β-PdBi$_2$. In fact, the Majorana zero mode has not been observed by the scanning tunneling microscopy measurements of β-PdBi$_2$ [24]. However, recent studies have reported that the SC transition temperature changes in thin films of β-PdBi$_2$, and there is a possibility that a different superconducting state emerges because surface states are more dominant for thin films [39-41]. The direct measurements of the SC gaps of thin films are desired in the future.

In summary, we demonstrated that the β-PdBi$_2$ could be a unique topological superconductor with a centrosymmetric crystal structure and distinct topological surface states at $E_F$ by laser ARPES. The SC gaps are isotropic for both bulk and surface states. The SC gap sizes between the bulk- and surface-derived bands differ about 0.1 meV, which further suggests that the proximity effect is particularly pronounced due to the bulk- and surface-derived bands lying close to each other in momentum space.

*Acknowledgments* - This work was supported by Grants-in-Aid for Scientific Research (KAKENHI) (Grants No. JP19H01818, No. JP19H00659, No. JP19H00651, No. JP 21H04652, No. JP21K18181, JP21H05235, No. JP24K01375, No. JP24K00565, No. JP24KF0021, and JP24K01285) from the Japan Society for the Promotion of Science (JSPS), by JSPS KAKENHI on Innovative Areas "Quantum Liquid Crystals" (Grant No. JP19H05826), and the Quantum Leap Flagship Program (Q-LEAP) (Grant No. JPMXS0118068681) from the Ministry of Education, Culture, Sports, Science, and Technology (MEXT). T.Su. acknowledges the research grants from the Izumi Science and Technology Foundation, Itoh Science Foundation, Toyota Riken Scholar, and Iketani Science and Technology Foundation. Y.Z. received support by a JSPS International Researcher Fellowship program.

Data availability—Data are available from the corresponding authors upon reasonable request.

*Contact author: suzuki.takeshi.fbs@osaka-u.ac.jp
†Present address: Graduate School of Frontier Biosciences, The University of Osaka, 1-3 Yamadaoka, Suita, Osaka 565-0871, Japan
‡Present address: National Metrology Institute of Japan, National Institute of Advanced Industrial Science and Technology, 1-1-1 Umezono, Tsukuba, Ibaraki 305-8563, Japan
§Present address: Graduate School of Informatics and Engineering, The University of Electro-Communications, Chofu, Tokyo 182-8585, Japan
¶Contact author: okazaki@issp.u-tokyo.ac.jp